\documentclass[aip,jap,superscriptaddress,reprint,twocolumn]{revtex4-1}

\usepackage{graphicx}
\usepackage{color}
\usepackage{amssymb}
\usepackage{amsmath}
\usepackage{bm}

\begin{document}

\title{Topological phase transition and quantum spin Hall state in TlBiS$_2$}

\author{Bahadur Singh}
\affiliation{Department of Physics, Indian Institute of Technology Kanpur, Kanpur 208016, India}

\author{Hsin Lin}
\affiliation{Graphene Research Centre and Department of Physics, National University of Singapore, Singapore 117542} 

\author{R. Prasad}
\affiliation{Department of Physics, Indian Institute of Technology Kanpur, Kanpur 208016, India}

\author{A. Bansil}
\affiliation{Department of Physics, Northeastern University, Boston, Massachusetts 02115, USA} 


\begin{abstract}
We have investigated the bulk and surface electronic structures and band topology of TlBiS$_2$ as a function of strain and electric field using \textit{ab-initio} calculations. 
In its pristine form, TlBiS$_2$ is a normal insulator, which does not support any non-trivial surface states. We show however that a compressive strain along the (111) direction 
induces a single band inversion with Z$_2$ = (1;000), resulting in a Dirac cone surface state with a large in-plane spin polarization. Our analysis shows that a critical point
lies between the normal and topological phases where the dispersion of the 3D bulk Dirac cone at the $\Gamma$-point becomes nearly linear. The band gap in thin films of TlBiS$_2$ 
can be tuned through an out-of-the-plane electric field to realize a topological phase transition from a trivial insulator to a quantum spin Hall state. An effective 
$\mathbf{k \cdot p}$ model Hamiltonian is presented to simulate our first-principles results on TlBiS$_2$.
\end{abstract}
 
\maketitle

\section{Introduction}

The interplay between spin-orbit coupling (SOC) and topological band theory gives rise to a new state of quantum matter known as a topological insulator (TI)\cite{hasan,qi,qshe}
in which band orderings become inverted in relation to their natural order at certain high symmetry points in the bulk Brillouin zone. TIs support spin-dependent conductive states at their boundaries while remaining 
insulating in the bulk, and are thus topologically distinct from the normal insulators. The topological surface states are protected by time reversal symmetry and exhibit nearly 
linear energy dispersion with unique spin-textures where spin is locked perpendicular to momentum. The topological protection guarantees backscattering-free transport at the
boundaries of TIs. Due to the existence of these novel surface states, the TIs not only offer potential applications in quantum computing and spintronics\cite{nayak_QC}, but
also pave the way for realizing novel quantum phenomena such as Weyl semimetals\cite{wan,murakami,singh1}, Majorana-fermions\cite{majorana} and Higgs mechanism\cite{higgs,mass}
in a condensed matter system. 

First principles calculations have led to the prediction of a large number of three-dimensional (3D) and two-dimensional (2D) TIs, which include
2D HgTe/CdTe quantum wells\cite{bervenig,konig}, bismuth and antimony based V$_2$VI$_3$ binary thermoelectrics\cite{zhang,hsieh,xia}, and many ternary compounds ranging from 
the Tl-compounds to the half-Heusler family, among others\cite{lin,yan_tl,neupane,heuslar,throughput,apervoskite}. TIs usually feature a band inversion, driven by the high SOC 
associated with the heavier atoms in the material, at the time reversal invariant momenta (TRIM) in the bulk Brillouin zone. Thus the tuning of SOC can provide a pathway for
inducing topological phase transitions (TPT) from the normal to the topological insulator state. One approach is to undertake chemical tuning, which however often
entails uncontrolled effects of chemical disorder. Another common approach is to attempt modifying the band topology by adjusting lattice constants or internal atomic positions
in the unit cell\cite{lin,singh1,heuslar,gst225_ahuja,hgse_blah,zhu_st,Inse,ab1,ab2,ab3}.

Many theoretical studies show that a 2D TI or a quantum spin Hall (QSH) insulator, along with TPTs therein, are possible by tuning the thickness of quantum wells or via reduced 
dimensionality of 3D TIs in the form of thin-films. \cite{lin,wada,liubi2te3,yzhangnat584,APL_Sb,singh2,InAsGaSbtheory,InAsGaSbexp}. The band topology of the thin films can also 
be controlled through an external transverse electric field\cite{APL_Sb,PNAS_sb2te3,PRL_hgcdte,APL_helical}, which could be provided by the substrate and/or controlled via gating.
Despite many theoretical proposals, the QSH state has been realized to date only in HgTe/CdTe and InAs/GaSb/AlSb quantum well systems\cite{InAsGaSbexp,konig}. The need for 
finding new 2D materials, in which the QSH state TPTs can be realized, is thus clear. 

In this paper, using systematic \textit{ab-inito} calculations, we show that a 3D topological insulating phase as well a TPT can be induced in TlBiS$_2$ through external strain,
and that thin films of this material can realize the QSH phase and undergo a TPT as a function of external electric field.  
TlBiS$_2$ is found to be a trivial insulator both theoretically and experimentally with a direct band gap at the $\Gamma$ point.\cite{lin,higgs,mass,singh1} It exhibits
a layered crystal structure with strong ionic and covalent type bonds within as well as between the layers. In sharp contrast, in the Bi$_2$Se$_3$ family of 
compounds,\cite{zhang,zhu_st,Inse} blocks of layers are held together by weak van der Waals type bonding.  
The strong bonding between the layers in TlBiS$_2$, on the other hand, makes this system a good prototype for the tuning and inversion of the band gap via the interlayer
distance.

Bearing the preceding considerations in mind, we delineate in this study the evolution of topological band order in TlBiS$_2$ with strain. Our analysis shows that a 3D-TPT 
occurs when a compressive strain is applied along the (111) direction, yielding an inverted band gap larger than 250 meV. The critical point between the normal and topological
phase is accompanied by the formation of a single 3D bulk Dirac cone at the bulk Brillouin zone center with a nearly linear energy dispersion over a substantial  energy range.
Our slab computations further show that the (111) surface of strained TlBiS$_2$ supports a single Dirac cone surface state at the $\Gamma$-point. This surface state displays a 
large in-plane spin-polarization with small hexagonal warping effects. Moreover, we show that a 2D TPT as well as the QSH state can be realized in thin films of pristine 
TlBiS$_2$ through an out-of-the-plane electric field. The critical point in this case supports six spin-polarized Dirac cones along the $\overline{\Gamma}-\overline{K}$ 
directions. Finally, we discuss a simple $\mathbf{k \cdot p}$ model Hamiltonian that captures the salient features of our first-principles bulk and surface states in TlBiS$_2$.   

The organization of the present article is as follows. In section II, we give the computational details and discuss the bulk crystal and band structures. Section III explains 
the evolution of topological band order in TlBiS$_2$ under strain. Section IV considers the surface electronic structures at various strain values and the associated non-trivial
spin-textures. The evolution of the QSH state through an out-of-the-plane electric field is discussed in Section V. In Sec. VI, we give the $\mathbf{k \cdot p}$ Hamiltonian for 
TlBiS$_2$. In Sec. IV, we make brief concluding remarks. 
  
\section{Computational methods and bulk band structures}
 
We employed the density functional theory (DFT) framework\cite{kohan} with projector augmented wave (PAW) method\cite{paw} as implemented in 
the VASP package\cite{vasp}. The generalized-gradient approximation (GGA)\cite{pbe} was used to model exchange-correlation effects. SOC was taken into account self-consistently
to treat relativistic effects. The relaxed structural parameters from Ref. ~\onlinecite{singh1} were used. The plane-wave cutoff energy of 350 eV was employed,
and an $8 \times 8 \times 8$ $\Gamma$-centered k-mesh was used for bulk computations. The strain was simulated by varying lattice constant '$c$' of TlBiS$_2$. 
For surface computations, we employed the slab model with periodically repeated slabs with a vacuum of 12 \r{A} and a $\Gamma$-centered  9 $\times$ 9 $\times$ 1 k-mesh. 
Surface relaxation is well-known to play an important role in slab computations\cite{lin,singh1}, and therefore, all atomic positions were optimized until the residual
force on each atom was less than 0.005 eV/\r{A}. In order to resolve spin-textures of topological states, we carried out computations on a fine k-mesh in the vicinity of
the Dirac point, and obtained the three spin components at each k-point from the expectation values of the three Pauli spin matrices $\sigma_x$, $\sigma_y$ and $\sigma_z$.

TlBiS$_2$ belongs to the Tl-family of compounds\cite{lin,yan_tl,singh1}. It has a rhombohedral crystal structure with four atoms in the primitive unit cell.
The space group is $D_{3d}^5$ ($R\bar{3}m$, SG no. 166). The conventional hexagonal unit cell has 12 atoms corresponding to three formula units in which 
the layers are stacked in the order Tl-S-Bi-S along the \textit{z}-axis, see Figure 1(a). Each Tl(Bi) layer is sandwiched between the two S layers, so that the bonding
between all the layers is strong\cite{singh1}, and the material is essentially 3D, and it is quite distinct from the Bi$_2$Se$_3$ family where the quintuple layer blocks
are bonded via weak van-der Waals forces. TlBiS$_2$ possesses inversion symmetry like Bi$_2$Se$_3$, with Tl or Bi layers providing inversion centers. As a result, topological
invariants for the band structure can be computed straightforwardly from the parities of the Bloch wave functions at the TRIM points\cite{kane}.

\begin{figure}[h!] 
\centering
  \includegraphics[width=0.5\textwidth]{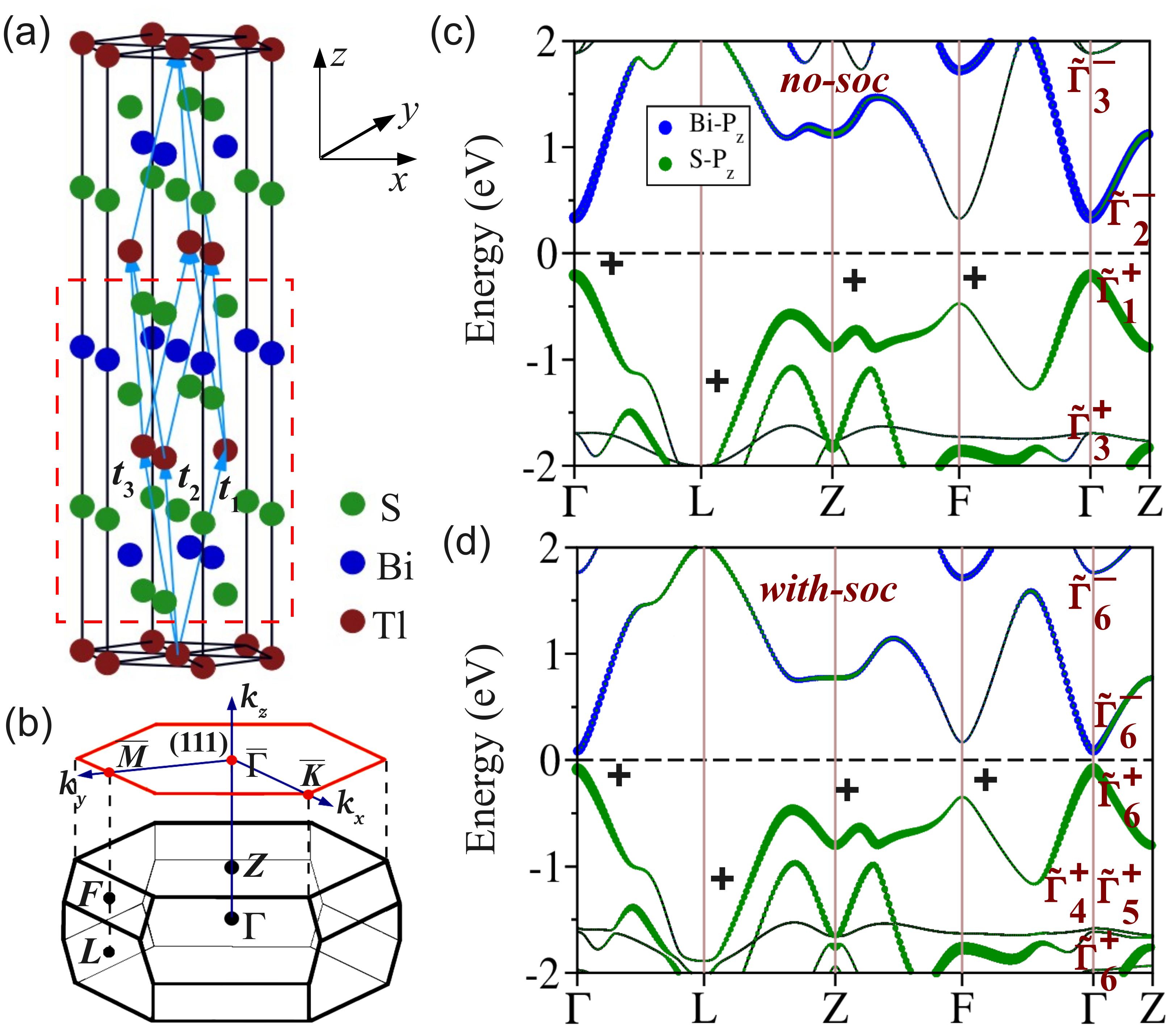}
  \caption {(a) Bulk hexagonal crystal structure of TlBiS$_2$ with layers stacked along the \textit{z}-axis in the order Tl-S-Bi-S. The primitive lattice vectors
  $\mathbf{t_{1,2,3}}$ are shown. 
	Red dotted box marks a block of seven atomic layers (7L). 
	(b) 3D-BZ corresponding to a primitive unit cell with four inequivalent TRIM points: $\Gamma, Z, F$, and $L$. 
	The 2D-BZ for the (111) surface with three high symmetry points: $\overline{\Gamma}, \overline{M}$ and $\overline{K}$. 
	Panels (c) and (d) show the bulk band structures without and with SOC, respectively. Sizes of blue and green dots represent contributions from Bi and S $p_z$ states
	to various bands. Irreducible representations of $D_{3d}^5$ group and its double group at the $\Gamma$-point are also shown in panels (c) and (d). Signs 
	of $\delta_i = \pm 1$  at the TRIM points are given\cite{footnote_delta}.}
 \end{figure} 
 
The bulk band structures without and with SOC are shown in Figs. 1(c) and (d). Without the SOC, TlBiS$_2$ is a direct gap semiconductor with a band gap of 0.64 eV at $\Gamma$. 
The band decomposed charge densities and symmetries show that the bulk valence band (BVB) is composed of S $p_z$ states and belongs to $\widetilde{\Gamma}^{+}_{1}$ representation
of space group $D_{3d}^5$\cite{zhang_model,yan_tl,footnote_parity}. In contrast, the bulk conduction band (BCB) is composed of Bi $p_z$ states and belongs
to $\widetilde{\Gamma}^{-}_{2}$ representation. A parity analysis\cite{kane} reveals that TlBiS$_2$ has Z$_2$ = (0;000). When we include the SOC, a large shift in the 
energies of all the bands occurs throughout the bulk Brillouin zone (BZ), and the band gap shrinks significantly to a value of 0.26 eV [Fig. 1(d)], but all bands remain 
two-fold spin degenerate. Within a spinor representation\cite{zhang_model}, the BVB and BCB belong to  $\widetilde{\Gamma}^{+}_{6}$ and $\widetilde{\Gamma}^{-}_{6}$ 
representations of the double group of $D_{3d}^5 ~(R\bar{3}m)$, respectively, as shown in Fig. 1(d). Although the SOC modifies the electronic structure significantly, it 
induces no band inversions, and the material remains a trivial insulator.

\section{3D Topological phase transition}

\subsection{Bulk analysis}

We have shown previously that the electronic structure of TlBi(S$_{1-x}$Se$_x$)$_2$ alloys \cite{singh1} can be tuned from a normal to a non-trivial insulating phase by modulating
the coupling potential. The inversion symmetry is explicitly broken in this case, yielding a semimetal phase at the critical point \cite{singh1}. Here, we consider the
ordered, inversion symmetric TlBiS$_2$ system and use a compressive strain to modify its electronic structure. Note in this connection that crystal field splitting (CFS) in 
TlBiS$_2$ lifts the degeneracy between the p$_z$ and p$_x$/p$_y$ orbitals\cite{yan_tl,zhang}. Any changes in the Bi$-$S and Tl$-$S bonds now lead to associated changes in the CFS, and hence in the band gap. Accordingly, we consider a uniaxial compressive strain along the trigonal axis and monitor the band gap for a possible band inversion. Relaxation of in-plane lattice constants '$a$' and '$b$' was found to have a negligible effect over the band gap at $\Gamma$ when interlayer distance '$c$' was reduced. For this reason,  
we fixed the in-plane lattice constants to their bulk relaxed values and varied only '$c$'. The strain is defined as $\delta c/{c_0}$ where $\delta c =c-c_0$, is the 
difference between the strained and relaxed lattice constant. 

\begin{figure}[ht!] 
\centering
\includegraphics[width=0.48\textwidth]{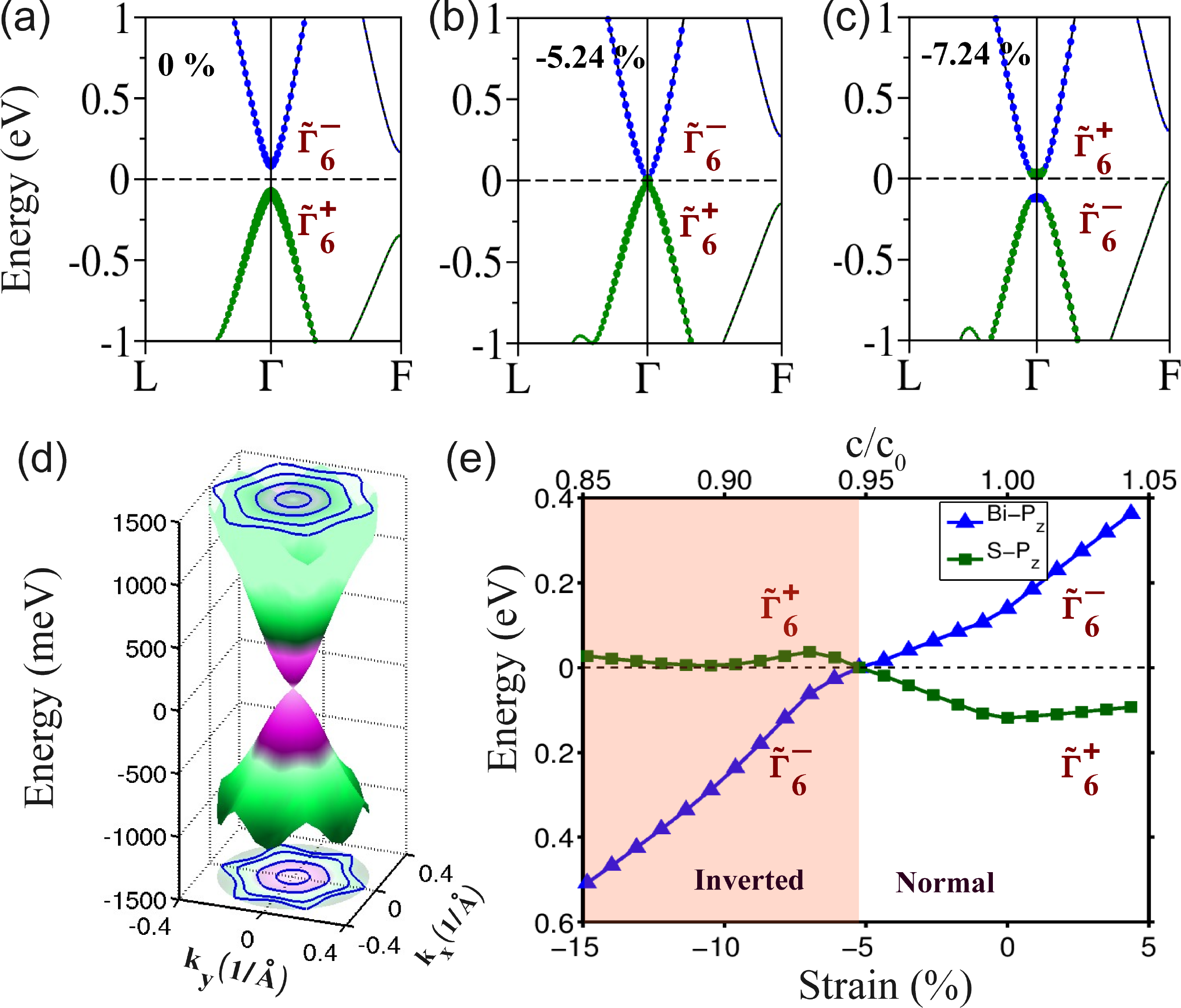}
\caption{ (a - c) Bulk band structure of TlBiS$_2$ with SOC along two high symmetry lines through the $\Gamma$-point at three different values of strain.
(d) A 3D bulk rendition of the band structure of panel (b) in the vicinity of the zone center $\Gamma$-point at the critical point. The constant energy contours for the BCB
and BVB are shown on the top and bottom planes. (e) Evolution of BVB and BCB at $\Gamma$. Shaded area gives the region where the band gap is inverted. Dashed (zero) line marks 
the Fermi energy.}
\end{figure}

The strain dependent band structure in Figs. 2(a)$-$(c) shows that as we increase the compressive strain along the (111) direction, the band gap starts decreasing as
the $\widetilde{\Gamma}^{+}_{6}$ BVB and $\widetilde{\Gamma}^{-}_{6}$ BCB levels start approaching each other, and closes at the critical strain value of -5.24\%, 
see Fig. 2(b). Based on the dependence of free energy on the lattice parameters\cite{murnaghan}, we estimate that a pressure of 3.1$-$3.5 GPa would be sufficient to 
induce such a strain. The band gap is seen from Fig. 2(c) to become inverted with a further increase in strain. An analysis of the band structure of Fig. 2(c) indicates
that the band gap extends over the entire BZ. Evolution of $\widetilde{\Gamma}^{+}_{6}$ and $\widetilde{\Gamma}^{-}_{6}$ levels over a wide range of strains is shown in
Fig. 2(e). The inverted band gap is seen to attain a value of 0.28 eV at -10.24\% strain, which is fairly large compared to other Tl compounds. It could be increased further
with increased strain, although such large strains would be difficult to realize in practice. We have verified the topological nature of band structure of TlBiS$_2$ at various 
strain values via parity analysis. Illustrative results at two different strain values, one at which the system is trivial and another at which it is non-trivial, are
presented in Table I. These results show that the topological phase is induced via a single band inversion at the $\Gamma$-point with increasing strain.

\begin{table}[h!] 
\centering
\caption{Products of parity eigenvalues at four TRIM points for 0~\%(trivial) and -7.24~\% (non-trivial) strain. The resulting $Z_2$ values are shown.}  
\begin{tabular}{ c  c   c   c  c   c }
\hline \hline
Strain    &  $\Gamma$  $\times$1  & L $\times$3 & F  $\times$3 & Z  $\times$1  &  Z$_2$  \\ 
\hline
 0~\%     &  $+$    & $+$   & $+$  & $+$ & (0;000) \\
-7.24~\%  &  $-$    & $+$   & $+$  & $+$ & (1;000) \\
\hline \hline
\end{tabular}
\label{paritytable}
\end{table}

Fig. 2(d) shows a 3D rendition of the Dirac cone states at the critical strain value of Fig. 2(b). Interestingly, the constant energy contours (CECs) are seen to be nearly
circular in shape up to about 800 meV with linear dispersion, and would allow access to Dirac physics over a fairly high energy range.

\subsection{Surface electronic structure and spin-texture }

Figs. 3(a)$-$(c) consider the surface electronic structure based on fully relaxed slabs of TlBiS$_2$ at three different values of strain. As expected, at 0\% strain 
(Fig. 3(a)), only a trivial band gap is seen without the presence of gapless surface states. With increasing strain, the gap between the bulk bands (shaded cyan region)
closes at the critical strain value in Fig. 3(b), and reopens with the appearance of Dirac cone states lying within the bulk band gap (Fig. 3(c)). The surface states in 
Fig. 3(c) cross the Fermi energy an odd number of times between the $\overline \Gamma$ and $\overline M$-points, demonstrating that strained TlBiS$_2$ is a strong topological 
insulator.  

\begin{figure}[ht!] 
\centering
\includegraphics[width=0.48\textwidth]{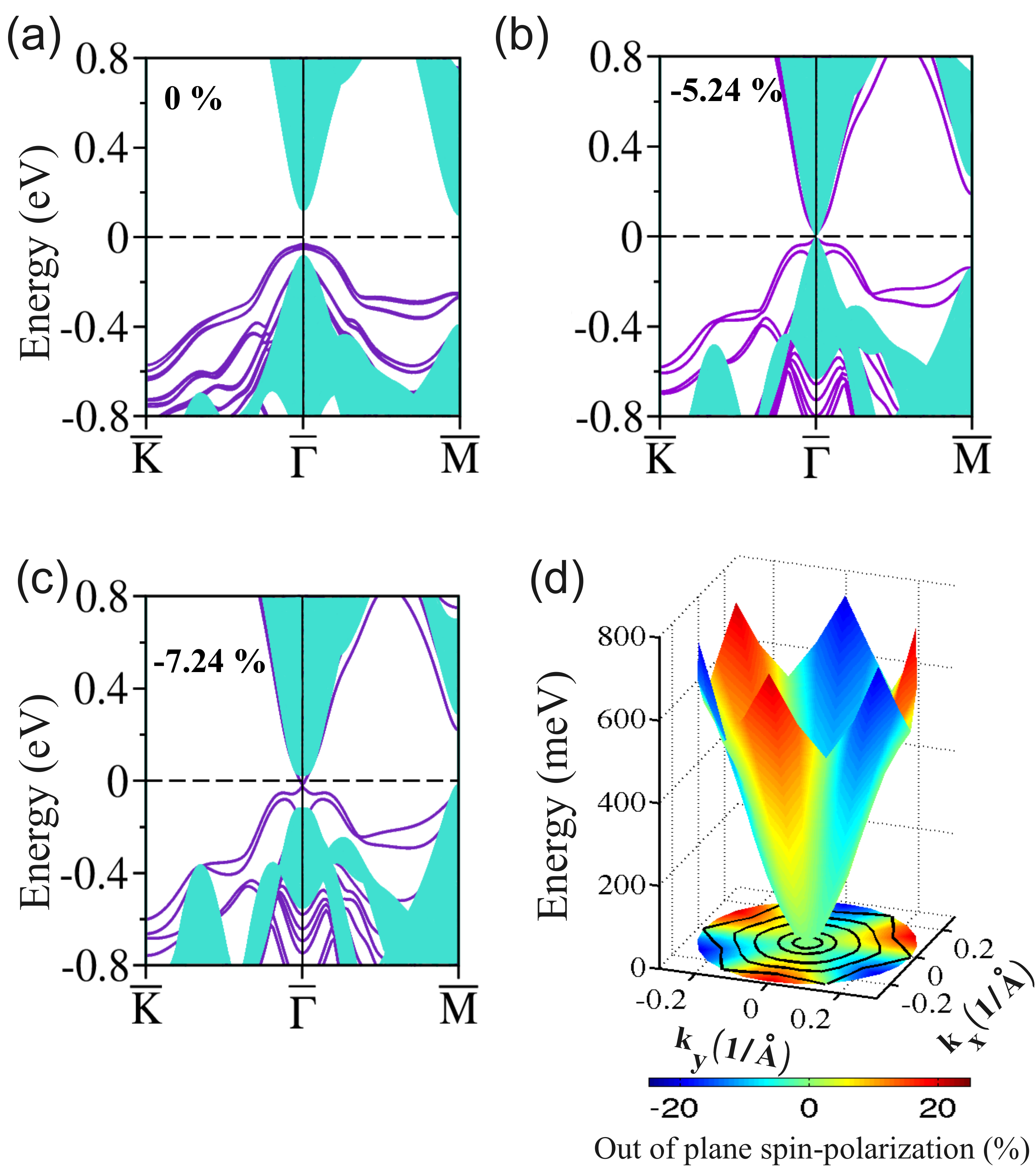}
\caption {(a)-(c) Surface electronic structure of TlBiS$_2$ at three different values of strain. The shaded cyan region shows the projected bulk bands.
(d) Energy dispersion of the upper Dirac cone surface state in the $k_z=0$ plane, along with the associated constant energy contours and the out-of-the-plane spin-polarization.} 
\end{figure}

Figure 3(d) shows the nearly linear 3D energy dispersion of the upper portion of the topological surface state in the $k_z=0$ plane with small hexagonal 
warping\cite{Hexawarp,susmita}. The CECs are essentially circular up to an energy of $\approx$ 500 meV. Spin textures of the upper and lower surface Dirac cones 
are further examined in Fig. 4. Spin is seen to be locked perpendicular to momentum over a large portion of the \textbf{k}-space with both the upper and lower cones
displaying distinct chiralities. As one moves away from the Dirac point, the spin-texture acquires a finite out-of-the-plane spin component as the Dirac cone becomes 
hexagonally warped.

\begin{figure}[h!] 
\centering
\includegraphics[width=0.49\textwidth]{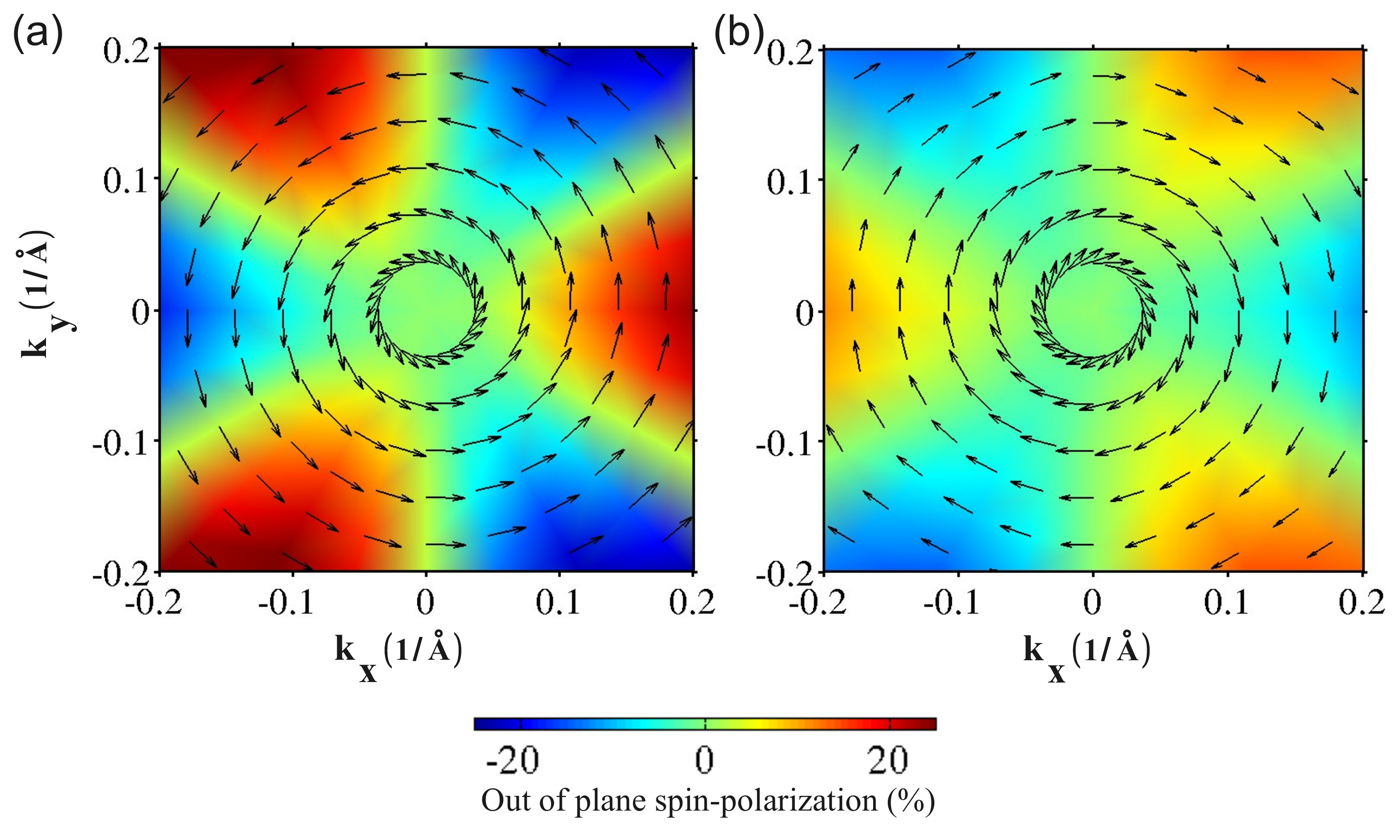}
\caption{In-plane spin direction of the non-trivial surface state of (a) lower and (b) upper Dirac cone. The out-of-the-plane spin component, shown by color bar, vanishes along
the mirror plane directions $\bar {\Gamma}-\bar{M}$.}
\end{figure}

\section{2D Topological phase transition and Quantum spin Hall state}

2D TPTs as well as the QSH state can be realized in thin 
films of 3D topological insulators through effects of reduced dimensionality \cite{liubi2te3,yzhangnat584,APL_Sb,singh2,wada}. In particular, when the film thickness becomes
smaller than the surface state decay length, quantum tunneling between the top and bottom surfaces of the film opens a thickness dependent gap at the Dirac point. The surface 
state decay length in Tl compounds is $\approx 3-5 nm$,\cite{lin,singh1} and therefore, films with less than about 30 atomic layers would be suitable for realizing the QSH state.
Strained films of TlBiS$_2$, which we have shown above to assume the 3D topological insulator phase, would thus be appropriate candidates for realizing a 2D TPT and
the QSH state\cite{singh2}. 

We will however explore an alternate route for achieving the QSH state in films of TlBiS$_2$ in the normal insulator phase by subjecting these films to an external electric
field (gating). \cite{APL_Sb,PNAS_sb2te3,PRL_hgcdte,APL_helical} Application of an out-of-the-plane electric field ($E_\perp$) breaks the inversion symmetry of the film as two 
sides of the film become inequivalent. As a result, the spin degeneracy of states away from the TRIM points is lifted. In this way, the electric field provides a mechanism for 
realizing spin-polarized states and gating controlled TPTs in thin films. Note that topological invariants now cannot be obtained through a parity analysis, but adiabatic 
continuity arguments must be deployed. We have specifically done so here by monitoring the band gap while varying strength of the SOC from 0-100\%. 

\begin{figure}[h!] 
\centering
\includegraphics[width=0.49\textwidth]{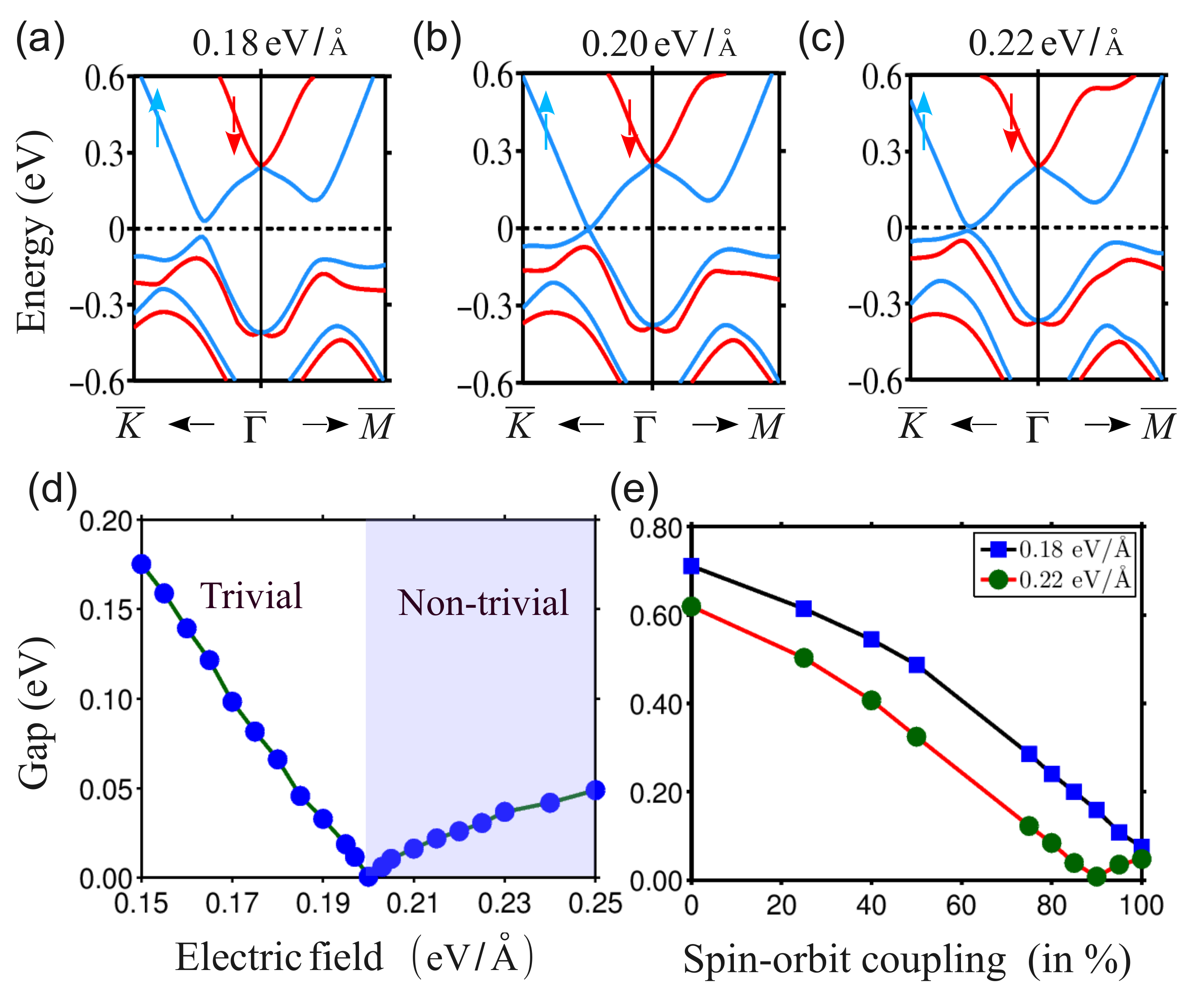}
\caption{Band structure of a seven layer (7L) TlBiS$_2$ slab under out-of-the-plane electric fields of: (a) 0.18 eV/\r{A}; (b) 0.20 eV/\r{A}; and, (c) 0.22 eV/\r{A}. 
The gap at the critical point [panel (b)] closes along the $\overline{\Gamma}-\overline{K}$ direction. Red and blue arrows represent the right and left handed spin helicities.
(d) Band gap along  $\overline{\Gamma}-\overline{K}$ as a function of electric field strength. The gap closes and reopen inverted beyond the critical field value of 0.20 eV/\r{A}.
Shaded area highlights the non-trivial field region. (e) Variation of band gap as a function of strength of the SOC at 0.18 eV/\r{A} and 0.22 eV/\r{A}.}
\end{figure}

The slab configuration with seven layers in which the S atom is in the top layer followed by Bi [See Fig. 1(a)] exhibits a minimum number of dangling bond 
states.\cite{lin,singh1} The band structure of such a relaxed 7L slab of TlBiS$_2$ is considered in Figs. 5(a$-$c) near the $\Gamma$-point under an external electric field,
$E_\perp$. In the absence of an electric field, the 7L thin film is an indirect band gap semiconductor with a band gap of 0.45 eV, and all states are at least twofold spin
degenerate due to inversion symmetry. As we apply the electric field, the spin degeneracy away from the TRIM points is lifted and the band gap begins to decrease. By monitoring
the band gap
size as a function of applied electric field, as shown in Figure 5(d), the gap closes at the critical field value of 0.20 eV/\r{A} along  the $\overline{\Gamma}-\overline{K}$
direction, see Figs. 5(b) and (d). At this critical point, the valence and conduction bands touch each other at six points, and form six spin-polarized Dirac cones along the 
$\overline{\Gamma}-\overline{K}$ directions. With further increase in the electric field the gap reopens (inverted) again the system assumes a QSH state; the topological nature
of this state can be established by monitoring the band gap as a function of SOC strength as shown in Fig. 5(e). For $E_\perp=0.18$~ eV/\r{A}, the gap decreases from 0.7 eV at 
0\%  to 0.08 eV at 100 \% SOC strength without closing at any intermediate value of the SOC. Therefore, the band structures with and without the SOC are connected adiabatically 
and hence are both topologically trivial.  On the other hand, for $E_\perp=0.22$ eV/\r{A}, the gap closes at 90 \% SOC strength and 
reopens with a value of 0.05 eV at 100 \%, allowing us to conclude that the 7L film is in the QSH state for fields greater than 0.20 eV/\r{A}. Along the preceding 
lines, we also investigated thicker slabs of 11L, 15L and 19L using atomic configuration similar to that of the 7L slab; all these slabs were found to exhibit the QSH state for
relatively small values of the external electric field. We also examined thinner 5L slabs with Bi or Tl atoms in the top layer. These slabs also transition into the QSH state
with critical field value of 0.10-0.15 eV/\r{A}, although the critical topological point now lies within the dangling bond states. The preceding results show clearly that
pristine TlBiS$_2$ films can be switched between the normal and QSH states under the action of a perpendicular external electric field.

\section{$\mathbf{k \cdot p}$ model Hamiltonian}

We now discuss a $\mathbf{k \cdot p}$ model Hamiltonian which captures the interesting physics of TlBiS$_2$. Since the bulk band structure undergoes band inversion at only 
the $\Gamma$-point, it is appropriate to consider an effective Hamiltonian near the $\Gamma$-point, which takes into account symmetry properties of the system\cite{Winkler_kp}.
For this purpose, we need to identify the irreducible representations of the symmetry group for both the valence and the conduction bands at the $\Gamma$-point. In the trivial
phase, the BVB is mainly composed of S $p_z$ anti-bonding states, whereas the BCB is composed of Bi $p_z$ bonding states. For convenience, we denote the BVB as $P2^+$ and BCB as 
$P1^-$, where $\pm$ indicates the parity of the corresponding states. The four states involved thus are: 
{($|P2^+,\uparrow \rangle$,$|P1^-,\uparrow \rangle$,$|P2^+,\downarrow \rangle$, $|P1^-,\downarrow \rangle$)}. The crystal structure of TlBiS$_2$ (rhombohedral with space group
$D_{3d}^5~(R\bar{3}m)$) is the same as that of Bi$_2$Se$_3$ and TlBiTe$_2$. The wavefunctions at $\Gamma$ can therefore be classified along the lines of Bi$_2$Se$_3$ and
TlBiTe$_2$\cite{zhang,zhang_model,yan_tl}, see Ref. ~\onlinecite{zhang_model} for details. As discussed in section II above, the BVB and BCB belong
to $\widetilde{\Gamma}^{+}_{6}$ and $\widetilde{\Gamma}^{-}_{6}$ representations, respectively. Keeping all this in mind, the four band effective Hamiltonian of Bi$_2$Se$_3$ 
is applicable here, which in the basis ($|P2^+,\uparrow \rangle$,$|P1^-,\uparrow \rangle$,$|P2^+,\downarrow \rangle$, $|P1^-,\downarrow \rangle$)\cite{zhang,zhang_model,yan_tl}, 
is:

\begin{equation}
 H(k)=H_0(k)+H_3(k)
\end{equation} 
\begin{equation}
 H_0(k)= \epsilon(k)\mathbb{I}_{4\times4} \\
  +\left(\begin{array}{cccc}
                               M(k)      &     -iB_0k_z      &      0           &  iA_0k_+       \\
	                    iB_0k_z    &     -M(k)            &  iA_0k_+   &     0                \\
                                  0       &     -iA_0k_-   &     M(k)         &   -iB_0k_z        \\
	                 -iA_0k_-   &        0             &  iB_0k_z      &    -M(k)             \\
\end{array}\right)
\end{equation}

\begin{equation}
\begin{split}
H_3(k)= \frac{R_1}{2} \left(\begin{array}{cccc}
                               0      &     K_+      &       0        &   0       \\
	                       K_+    &      0        &       0       &  0        \\
	                       0       &      0        &      0       &  -K_+        \\
	                      0       &      0        &      -K_+     &  0           \\
\end{array}\right) \\ 
+ \frac{R_2}{2}      \left(\begin{array}{cccc}
                               0      &     -K_-      &       0        &  0       \\
	                       K_-    &      0        &       0        &  0        \\
	                       0       &      0        &      0       &  -K_-       \\
	                      0       &      0        &      K_-       &  0          \\
\end{array}\right) 
\end{split}
\end{equation}

with $k_\pm=k_x\pm ik_y$, $\epsilon(k) = C_0 +C_1k_z^2+C_2(k_x^2+k_y^2)$, $M(k)=M_0 - M_1k_z^2 -M_2(k_x^2+k_y^2)$, and $K_\pm=k_+^3 \pm k_-^3$. Here, $H_0(k)$ preserves the 
in-plane rotation symmetry, whereas $H_3(k)$ reduces it to a threefold rotation symmetry. The parameters $C_0$, $C_1$, $C_2$, $M_0$, $M_1$, $M_3$, $A_0$, $B_0$, $R_1$, and $R_2$ 
are calculated numerically by fitting to \textit{ab-inito} dispersions, and are given for 0\% (trivial region)  and -7.24 \% (non-trivial region) strain in 
Table~\ref{param_model}. At -7.24 \% strain, $M_0 > 0$, $M_1 > 0$, and $M_2 > 0$, which implies that the system stays in the inverted region and thus remains topologically
non-trivial for larger strains.  

\begin{table}[h!] 
\centering
\caption{Parameters for the model Hamiltonian of Eq. (2) at 0 \% (trivial) and -7.24 \% (non-trivial) strain. } 
\begin{tabular}{l c c c}
\hline \hline
  
  Parameter             &     0\%         &   -7.24\%    \\
\hline
$A_0$ (eV \r{A})         &    1.04         &    1.83      \\
$B_0$ (eV \r{A})         &    3.06         &    2.72      \\
$C_0$ (eV)               &   -0.0035       &  -0.0011     \\
$C_1$ (eV \r{A}$^{2}$)   &    1.97         &    4.3       \\
$C_2$ (eV \r{A}$^{2}$)   &    3.616	    &    2.337     \\
$M_0$ (eV )              &    -0.0859      &   0.0714     \\
$M_1$ (eV \r{A}$^{2}$)   &    2.14         &   0.368      \\
$M_2$ (eV \r{A}$^{2}$)   &    21.08        &   18.67      \\
$R_1$ (eV \r{A}$^{3}$)   &    22.81        &   16.12      \\
$R_2$ (eV \r{A}$^{3}$)   &    48.23        &   38.23      \\

\hline \hline
\end{tabular}
\label{param_model}
\end{table}

The non-trivial surface dispersion for TlBiS$_2$ can be directly computed from the above $4 \times 4$ model Hamiltonian using appropriate boundary
conditions\cite{konig,zhang_model,yan_tl}. Here, we consider the semi-infinite system for which the Hamiltonian of Eq. (1) applies only for $z>0$. It can then be 
shown straightforwardly that the two localized states $|\psi_\uparrow\rangle$ and $|\psi_\downarrow\rangle$, which are Kramer’s partners, appear at the surface only
if $M_0M_1>0$, i.e. in the inverted regime. The effective Hamiltonian for the surface can be extracted by projecting the bulk Hamiltonian (1) on to the space spanned
by these two states as:

\begin{equation}
\begin{split}
 H_{surf}(k)=\tilde{C_0}+\tilde{C_2}(k_x^2+k_y^2)+\tilde{A_0}(k_x\sigma_y-k_y\sigma_x) \\
 + \tilde{R_1\over 2} (k_+^3+k_-^3)\sigma_z
\end{split}
\end{equation} 

Here, the parameters $\tilde{C_0}$, $\tilde{C_2}$, $\tilde{A_2}$, and $\tilde{R_1}$ depend upon the material and the boundary conditions. The $k^3$ term is coupled to
$\sigma_z$ and breaks the in-plane rotation symmetry to a three fold rotation symmetry. This is the counterpart of cubic Dresselhaus spin-orbit coupling in the bulk
rhombohedral structures and ensures the appearance of an out-of-the-plane spin-component, and explains the hexagonal warping observed in Figure 5 \cite{Hexawarp}. 
The $k^3$ term, and hence the $S_z$ component, vanishes along the $\overline \Gamma-\overline M$ mirror directions. 

\begin{figure}[h!] 
\centering
\includegraphics[width=0.4\textwidth]{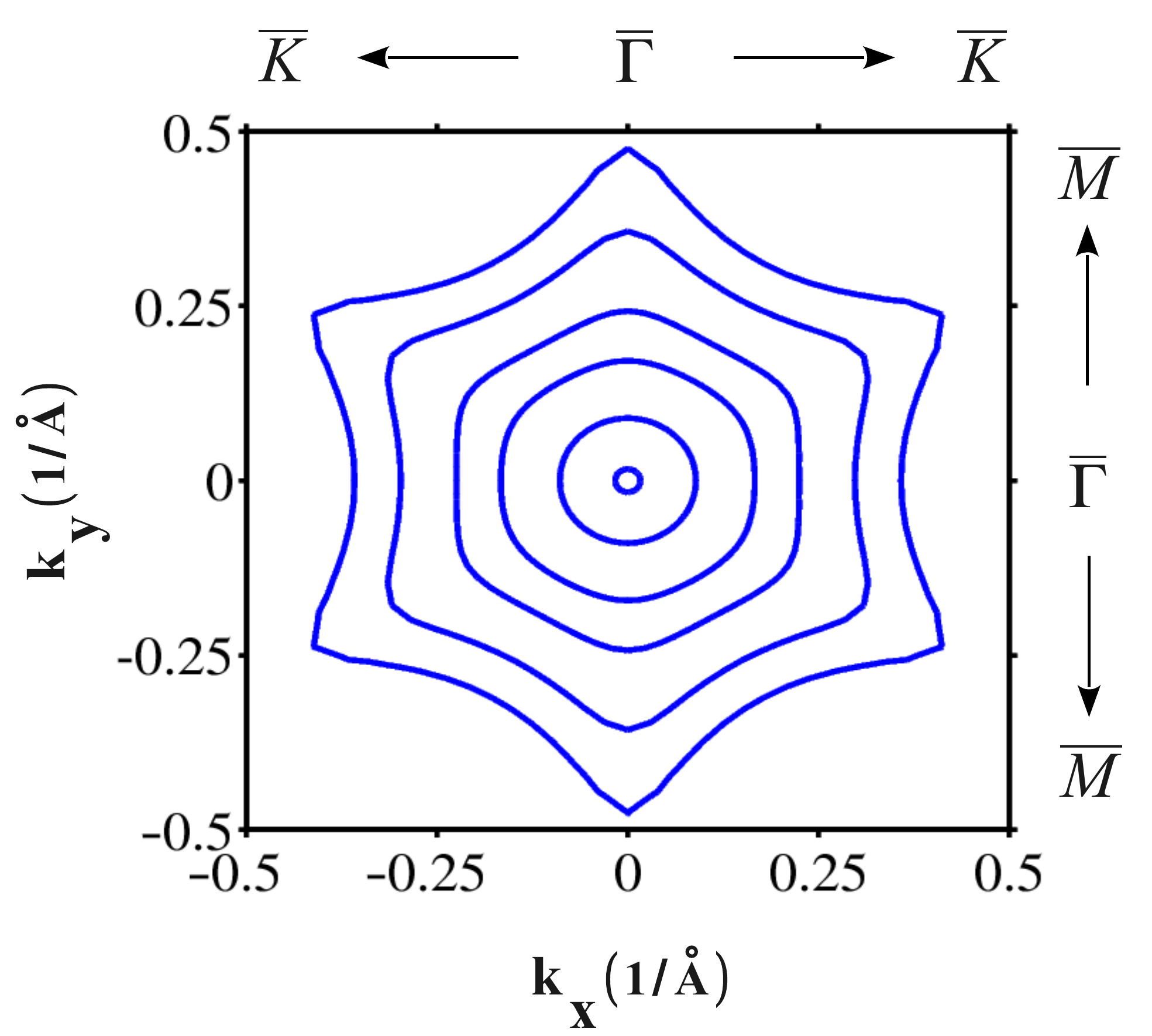}
\caption{Constant energy contours of $H_{surf}(k)$ displaying six fold symmetry where vertices of the hexagon lie along the $\overline \Gamma-\overline M$ directions. }
\end{figure}

The surface band dispersion of $H_{surf}(k)$ using $k_\pm=k_x\pm ik_y=ke^{\pm\theta}$, and the bulk values of the parameters in equation (4), is:
\begin{equation}
 E_{\pm}(k)=C_0+C_2k^2\pm \sqrt{A_0^2 k^2+R_1^2 k^6 \cos^2(3\theta)} 
\end{equation}
where $E_{\pm}$ denotes the energy of upper (conduction) and lower (valence) bands. Figure 6 shows the CECs for upper (conduction) band calculated using this dispersion. 
The CECs are circular near the Dirac point, but as we move away from the Dirac point, the CECs become hexagonal due to $k^3$-term \cite{Hexawarp}. The hexagonal warping is
maximum along the $\overline \Gamma-\overline K$ directions, whereas it vanishes along the $\overline \Gamma-\overline M$ mirror directions. As we further increase the energy
away from the Dirac point, the CECs attain a snowflakes shape where vertices lie along the $\overline \Gamma-\overline M$ directions. These features of CECs are in good agreement
with our \textit{ab-initio} results.

\section{Conclusion}
We discuss how strain and electric field could be used to tune the bulk and surface electronic structures of TlBiS$_2$ using \textit{ab-intio} DFT calculations. 
In its relaxed, pristine structure TlBiS$_2$ is found to be a normal insulator without any surface states.  However, the system undergoes a 3D topological phase transition
under 5.24 \% compressive strain with the formation of 3D bulk Dirac cone states with nearly linear energy dispersion. The inverted band gap attains a value of 0.28 eV for 10.24\% 
compressive strain, and the band gap could be increased further by increasing strain. Through slab computations, we show that Dirac cone surface state has a nearly linear 
energy dispersion with a large in-plane spin-polarization. We show that the a 2D topological phase transition from a normal insulator to the QSH state can be realized in thin
films of TlBiS$_2$ by varying the strength of an electric field perpendicular to the film surface. Finally, we present a $\mathbf{k \cdot p}$ model Hamiltonian for bulk and 
surface states of strained TlBiS$_2$. Our results indicate that TlBiS$_2$ films are a viable candidate for realizing a gating controlled on/off switch between the normal and 
QSH states.

\section*{Acknowledgments} 
This work was supported by the Department of Science and Technology, New Delhi (India) through project SR/S2/CMP-0098/2010. H.L. acknowledge 
the Singapore National Research Foundation for support under NRF Award No. NRF-NRFF2013-03. The work at Northeastern University is supported by
the US Department of Energy, Office of Science, Basic Energy Sciences contract DE-FG02-07ER46352, and benefited from theory support at the Advanced 
Light Source and the allocation of supercomputer time at NERSC through DOE grant number DE-AC02-05CH11231.


\begin{thebibliography}{10}
\bibitem{ab1} It will be interesting to consider combined effects of strain and disorder on the surface and bulk electronic structures using various approaches;  see, for example, Refs. \onlinecite{ab2,ab3}.
\bibitem{footnote_delta}
$\delta_i$ represents the product of parity eigenvalues of the occupied bulk bands at the TRIM points and is given as
$  \delta_i=\prod_{m=1}^N \xi_{2m}(\Gamma_i),$ where N is the number of occupied bulk bands and $\xi_{2m}(\Gamma_i)=\pm{1}$ is the parity eigenvalue of the 2m$^{th}$ occupied 
energy band at the point $\Gamma_i$.
\end{thebibliography}
\end{document}